\documentclass[]{piparticle-final}
\usepackage{graphicx}
\usepackage{amssymb}
\usepackage{amsmath}
\usepackage{cite}

\begin{document}
\volume{2}               
\articlenumber{020008}   
\journalyear{2010}       
\editor{A. Vindigni}   
\reviewers{A. A. Fedorenko, CNRS-Lab. de Physique, ENS de Lyon, France.}  
\received{20 October 2010}     
\accepted{1 December 2010}   
\runningauthor{S. Bustingorry \itshape{et al.}}  
\doi{020008}         

\title{Anisotropic finite-size scaling of an elastic string at the depinning threshold in a random-periodic medium }

\author{S. Bustingorry,\cite{inst1}\thanks{E-mail: sbusting@cab.cnea.gov.ar}
        A. B. Kolton\cite{inst1}\thanks{E-mail: koltona@cab.cnea.gov.ar}
	}

\pipabstract{
We numerically study the geometry of a driven elastic string at its sample-dependent depinning threshold in random-periodic media. We find that the anisotropic finite-size scaling of the average square width $\overline{w^2}$ and of its associated probability distribution are both controlled by the ratio $k=M/L^{\zeta_{\mathrm{dep}}}$, where $\zeta_{\mathrm{dep}}$ is the random-manifold depinning roughness exponent, $L$ is the longitudinal size of the string and $M$ the transverse periodicity of the random medium. The rescaled average square width $\overline{w^2}/L^{2\zeta_{\mathrm{dep}}}$ displays a non-trivial single minimum for a finite value of $k$. We show that the initial decrease for small $k$ reflects the crossover at $k \sim 1$ from the random-periodic to the random-manifold roughness. The increase for very large $k$ implies that the increasingly rare critical configurations, accompanying the crossover to Gumbel critical-force statistics, display anomalous roughness properties: a transverse-periodicity scaling in spite that $\overline{w^2} \ll M$, and subleading corrections to the standard random-manifold longitudinal-size scaling. Our results are relevant to understanding the dimensional crossover from interface to particle depinning.
}
\maketitle

\blfootnote{
\begin{theaffiliation}{99}
   \institution{inst1} CONICET, Centro At{\'{o}}mico Bariloche, 8400 San Carlos
de Bariloche, R\'{\i}o Negro, Argentina.
\end{theaffiliation}
}

\section{Introduction}
\label{sec:intro}

The study of the static and dynamic properties of $d$-dimensional elastic
interfaces in 
$d+1$-dimensional random media is
of interest in a wide range of physical systems. Some concrete experimental
examples are
magnetic~\cite{lemerle_domainwall_creep,bauer_deroughening_magnetic2,
yamanouchi_creep_ferromagnetic_semiconductor2,metaxas_depinning_thermal_rounding} 
or ferroelectric~\cite{paruch_ferro_roughness_dipolar,paruch_ferro_quench}
domain walls, contact lines of
liquids~\cite{moulinet_distribution_width_contact_line2},
fluid invasion in porous media~\cite{Martys1991,Hecht2004}, and
fractures~\cite{bouchaud_crack_propagation2,alava_review_cracks}.
In all these systems, the basic physics is controlled by the competition between
quenched disorder 
(induced by the presence of impurities in the host materials) which promotes the 
wandering of the elastic object, against the elastic forces which tend to make
the elastic
object flat. One of the most dramatic and worth understanding manifestations 
of this competition is the response of these systems to an external drive.

The mean square width or roughness of the interface is one of the most basic 
quantities in the study of pinned interfaces. In the absence of an 
external drive, the ground state of the system is disordered but 
well characterized by a self-affine rough geometry with a diverging 
typical width $w \sim L^{\zeta_{\mathrm{eq}}}$, where $L$ 
is the linear size of the elastic object and $\zeta_{\mathrm{eq}}$ is the 
equilibrium roughness exponent. When the external force is increased from zero, 
the ground state becomes unstable and the interface is locked in metastable
states. 
To overcome the barriers separating them and reach a 
finite steady-state velocity $v$ it is necessary to exceed a 
finite critical force, above which barriers disappear and no 
metastable states exist. For directed $d$-dimensional elastic interfaces 
with convex elastic energies in a $D=d+1$ dimensional space with disorder, the 
critical point is unique, characterized by the 
critical force $F = F_c$ and its associated critical 
configuration~\cite{middleton_theorem}. 
This critical configuration is also rough and 
self-affine such that
$w \sim L^{\zeta_{\mathrm{dep}}}$ with $\zeta_{\mathrm{dep}}$ the depinning
roughness exponent. When approaching the threshold from above, the steady-state
average 
velocity vanishes like $v \sim (F-F_c)^\beta$ and the correlation length
characterizing
the cooperative avalanche-like motion diverges as $\xi \sim
(F-F_c)^{-\nu}$ for $F > F_c$, where $\beta$ is the velocity exponent and  $\nu$
is the depinning correlation length
exponent~\cite{fisher_depinning_meanfield,narayan_fisher_cdw,
nattermann_stepanow_depinning,ledoussal_frg_twoloops}. At finite temperature
and for $F \ll F_c$, the system presents an ultra-slow steady-state creep
motion with universal
features~\cite{ioffe_creep,chauve_creep} directly correlated with its 
multi-affine geometry~\cite{kolton_creep2,
kolton_creep_exact_pathways}. 
At very small temperatures the absence of a divergent correlation length below
$F_c$ shows
that depinning must be regarded as a non-standard phase 
transition~\cite{kolton_creep_exact_pathways,kolton_depinning_zerot2} while 
exactly at $F=F_c$, the transition is smeared-out with  
the velocity vanishing as $v \sim T^\psi$, with $\psi$, the so-called thermal
rounding
exponent~\cite{chen_marchetti,vandembroucq_thermal_rounding_extremal_model,
nowak_thermal_rounding,roters_thermal_rounding1,
bustingorry_thermal_rounding_epl,bustingorry_thermal_rounding_exponent}.

During the last years, numerical simulations have played an important role
to understand the physics behind the depinning transition thanks to the
development 
of powerful exact algorithms. In particular, the development of an exact
algorithm able to target efficiently the critical configuration and critical 
force for a given sample~\cite{rosso_roughness_MC,rosso_anharmonic_kpz} 
has allowed to study, precisely, the self-affine rough geometry at 
depinning~\cite{
moulinet_distribution_width_contact_line2,rosso_roughness_MC,rosso_anharmonic_kpz,
rosso_depinning_longrange_elasticity,rosso_width_distribution}, the sample-to-sample 
critical force distribution~\cite{bolech_critical_force_distribution}, 
the critical exponents of the depinning
transition~\cite{bustingorry_thermal_rounding_epl,bustingorry_thermal_rounding_exponent,duemmer2}, the renormalized disorder 
correlator~\cite{rosso_numerical_depinning_correlator}, 
and the avalanche-size distribution in quasistatic motion~\cite{rosso_avalanche_FRG}.
Moreover, the same algorithm has allowed to study, precisely, the transient
universal dynamics at depinning~\cite{kolton_short_time_exponents,kolton_universal_aging_at_fc}, 
and an extension of it has allowed to study low-temperature creep dynamics  
~\cite{kolton_creep_exact_pathways,kolton_depinning_zerot2}.

\begin{figure}[!tbp]
\begin{center}
\includegraphics[width=0.45\textwidth]{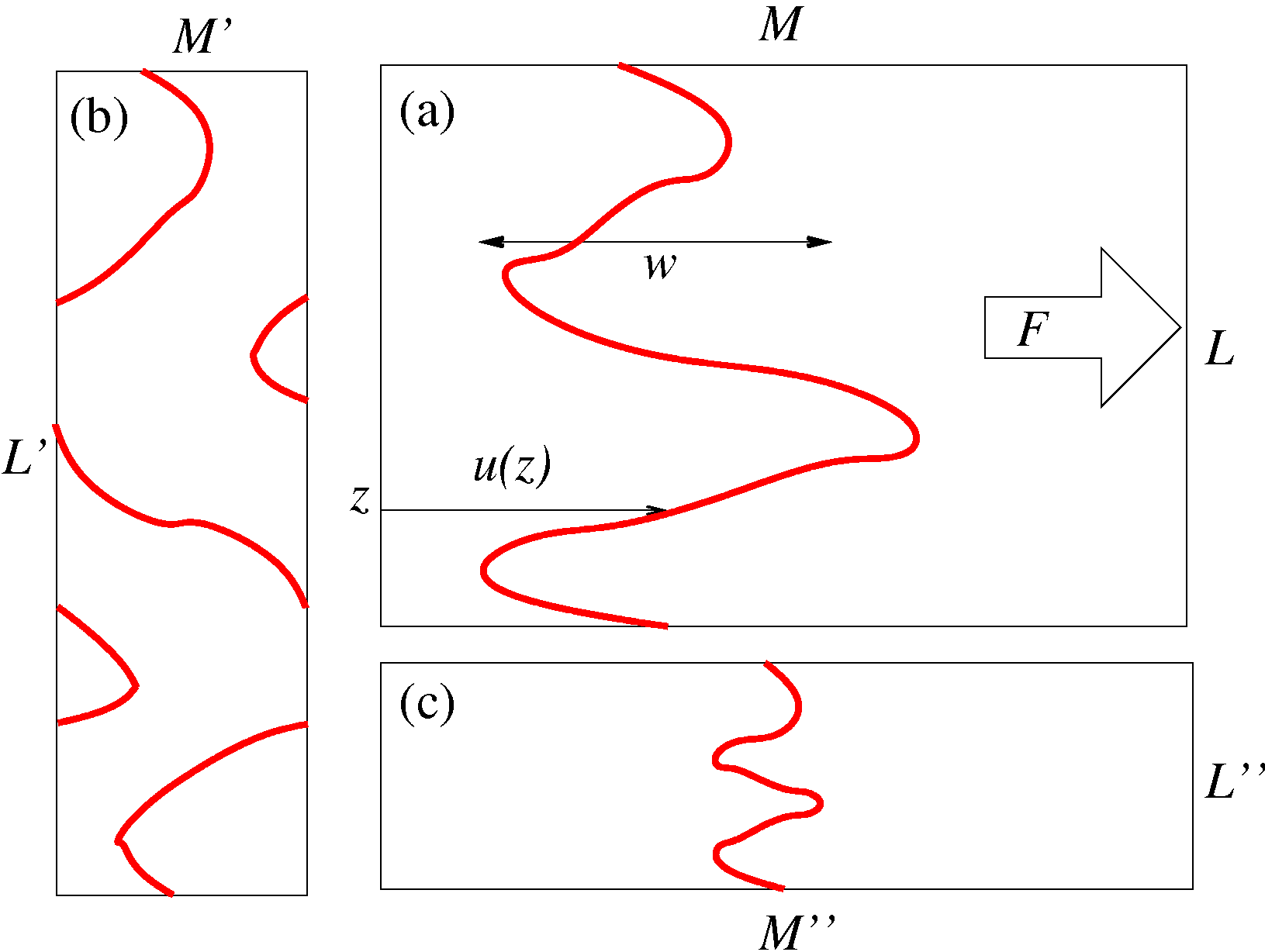}
\end{center}
\caption{\label{fig:schema}
(a) Elastic string driven by a force $F$ in a random-periodic medium 
with periodic boundary conditions. It is described by a displacement field $u(z)$ and 
has a mean width $w$. 
The anisotropic finite-size scaling of width fluctuations are controlled by the
aspect-ratio parameter $k=M/L^{\zeta_{\mathrm{dep}}}$, with ${\zeta_{\mathrm{dep}}}$ 
the random-manifold roughness exponent at depinning. 
In the case $k \ll 1$ (b) periodicity effects are important, while when $k \gg 1$ 
(c) they are not important but the roughness scaling of the critical configuration 
is anomalous.}
\end{figure}

In practice, the algorithm for targeting the critical configuration
~\cite{rosso_roughness_MC,rosso_anharmonic_kpz} has been numerically 
applied to directed interfaces of linear size $L$ displacing 
in a disordered potential of transverse dimension $M$, applying 
periodic boundary conditions in both directions in order to avoid  
border effects. This is thus equivalent to an elastic string displacing 
in a disordered cylinder. The aspect ratio between longitudinal $L$ and transverse $M$ 
periodicities must be carefully chosen, in order to have the desired 
thermodynamic limit corresponding to a given experimental realization. In Ref.~\cite{bolech_critical_force_distribution} it was indeed shown that
the critical force distribution $P(F_c)$ displays three regimes
associated with $M$: (i)
At very small $M$ compared with the typical width $L^{\zeta_{\mathrm{dep}}}$ 
of the interface, the interface 
wraps the computational box several times in the transverse direction, as shown schematically 
in Fig.~\ref{fig:schema}(b), and therefore the periodicity of the random medium is 
relevant and $P(F_c)$ is Gaussian; (ii)
At very large $M$ compared with $L^{\zeta_{\mathrm{dep}}}$, 
as shown schematically 
in Fig.~\ref{fig:schema}(c),  
periodicity effects are absent but then the critical force, being the maximum 
among many independent sub-critical forces, obeys extreme
value statistics and $P(F_c)$ becomes a Gumbel distribution; 
(iii) In the intermediate regime, where
$M \approx  L^{\zeta_{\mathrm{dep}}}$ and periodicity effects are still
irrelevant, as shown schematically 
in Fig.~\ref{fig:schema}(a), the distribution function is in between 
the Gaussian and the Gumbel distribution. It has
been argued that only the last 
case, where $M \approx L^{\zeta_{\mathrm{dep}}}$, corresponds to the
random-manifold depinning universality class (periodicity effects absent) 
with a finite critical force in the thermodynamic limit $L,M \to \infty$. 
This criterion does not give, however, the optimal value of the proportionality 
factor between $M$ and $L^{\zeta_{\mathrm{dep}}}$, and
must be modified at finite velocity 
since the crossover to the random-periodic universality class at large
length-scales depends also on the velocity~\cite{bustingorry_RM-RP}. 
To avoid this problem, it has been therefore proposed to define the critical scaling in the 
fixed center of mass ensemble~\cite{fedorenko_frg_fc_fluctuations}.
The crossover from the random-manifold to the random-periodic universality class 
is, however, physically interesting, as it can occur in periodic elastic systems
such as elastic chains. Remarkably, although the mapping
from a periodic elastic system (with given lattice parameter) in a random potential to a non-periodic elastic
system (such as an interface) in a random potential with periodic boundary conditions is not exact, it was recently shown that the lattice parameter does play 
the role of $M$ for elastic interfaces with regard to the geometrical or roughness
properties~\cite{bustingorry_RM-RP}. 
Since the periodicity can often be experimentally tuned in such periodic systems it is thus worth
studying in detail the geometry of critical interfaces of size $L$ as a function of $M$ with
periodic boundary conditions, and thus complement the 
study of the critical force in such systems~\cite{bolech_critical_force_distribution}.

In this paper, we study in detail, using numerical simulations, the geometrical 
properties of the one-dimensional interface 
or elastic string critical configuration in a 
random-periodic pinning potential as a function of the 
aspect ratio parameter $k$, conveniently defined as
$k=M/L^{\zeta_{\mathrm{dep}}}$.
We show that $k$ is indeed the only parameter controlling the finite-size
scaling (i.e. the dependence of observables with the dimensions $L$ and $M$) of the 
average square width and its 
sample-to-sample probability distribution.
The scaled average square width $\overline{w^2}L^{-2{\zeta_{\mathrm{dep}}}}$ 
is described by a universal function of $k$ 
displaying a non-trivial single minimum at a finite value of $k$. 
We show that while for small $k$ this reflects the 
crossover at $k \sim 1$ from the random-periodic to the random-manifold 
depinning universality class, for large $k$ it implies that 
in the regime where the depinning threshold is controlled by extreme 
value (Gumbel) statistics, critical configurations also become rougher, 
and display an anomalous roughness scaling.

\section{Method}
\label{s:model}

The model we consider here is an elastic string in $(1+1)$ dimensions described by a
single valued function $u(z,t)$, which gives the transverse displacement $u$ as
a function of the longitudinal direction $z$ and the time $t$ [see Fig.~\ref{fig:schema}(a)]. 
The zero-temperature dynamics of the model is given by
\begin{equation}
\label{eq:EW-F}
\gamma \, \partial_t u(z,t) = c \, \partial^2_z u(z,t) + F_p(u,z) + F,
\end{equation}
where $\gamma$ is the friction coefficient and $c$ the elastic constant. 
The first term in the right hand side derives from an harmonic elastic energy. 
The effects of a random-bond type disorder is given by the pinning force
$F_p(u,z) = - \partial_u U(u,z)$. The disorder potential $U(u,z)$ has zero
average and sample-to-sample fluctuations given by
\begin{equation}
\label{eq:Ucorr}
\overline{\left[ U(u,z)- U(u',z') \right]^2} = \delta(z-z') \, R^2(u-u'),
\end{equation}
where the overline indicates average over disorder realizations and $R(u)$
stands for a correlator of finite range $r_f$~\cite{chauve_creep}. Finally, $F$
represents the uniform external drive acting on the string. Physically,  
this model can phenomenologically describe, for instance, a magnetic domain wall in 
a thin film ferromagnetic material with weak and randomly located imperfections
~\cite{lemerle_domainwall_creep}, being $F$ proportional to an applied 
external magnetic field pushing the wall in the energetically favorable direction.

In order to numerically solve Eq.~\eqref{eq:EW-F}, the system is discretized in
the $z$-direction in $L$ segments of size $\delta z=1$, i.e. $z \to
j=0,...,L-1$, while keeping $u_j(t)$ as a continuous variable. To model the
continuous random potential, a cubic spline is used, which passes through $M$
regularly spaced uncorrelated Gaussian number
points~\cite{rosso_depinning_longrange_elasticity}. For the numerical simulations
performed here we have used, without loss of generality, 
$\gamma=1$, $c=1$ and $r_f=1$ and a disorder intensity $R(0)=1$. 
In both spatial dimensions we have
used periodic boundary conditions, thus defining a $L \times M$ system.

The critical configuration $u_c(z)$ and force $F_c$ are defined 
from the pinned (zero-velocity) configuration with the largest driving force $F$
in the long time limit dynamics.
They are thus the real solutions of
\begin{equation}
\label{eq:EW-F-root}
c \, \partial^2_z u(z) + F_p(u,z) + F = 0,
\end{equation}
such that for $F>F_c$ there are no further real solutions (pinned configurations). 
Middleton theorems~\cite{middleton_theorem} assure that for Eqs.~(\ref{eq:EW-F-root}) the solution exists 
and it is unique for both $u_c(z)$ and $F_c$, 
and that above $F_c$ the string trajectory in an $L$ dimensional phase-space 
is trapped into a periodic attractor (for a system 
with periodic boundary conditions as the one we consider). In other words, the critical 
configuration is the marginal 
fixed point solution or critical state of the dynamics, being $F_c$ the 
critical point control parameter of a Hopf bifurcation. 
Solving the $L$-dimensional system of Eqs.~(\ref{eq:EW-F-root}) for large $L$ directly is a formidable task, due to the  
non-linearity of the pinning force $F_p$. On the other hand, solving the long-time dynamics at different driving forces $F$ 
to localize $F_c$ and $u_c$ is very inefficient due to the critical slowing down. Fortunately, 
Middleton theorems, and in particular the ``non-passing rule'', can be used again to devise a 
precise and very efficient algorithm 
which allows to obtain the critical force $F_c$ and the critical configuration $u^c_j$ 
for each independent disorder realization iteratively without solving the actual dynamics nor 
directly inverting the system of Eqs.~(\ref{eq:EW-F-root})~\cite{rosso_depinning_longrange_elasticity}. 
Once the critical force and the
critical configuration are determined with this algorithm, we can compute the different observables. 
In particular, the square width or
roughness of the string at the critical point for a given disorder realization
is defined as
\begin{equation}
\label{eq:def-roughness}
w^2 = \frac{1}{L} \sum_{j=0}^{L-1} \left[ u^c_j - \frac{1}{L} \sum_{k=0}^{L-1}
u^c_k \right]^2.
\end{equation}
Computing $w^2$ for different disorder realizations allows us to compute
its disorder average $\overline{w^2}$ and the sample-to-sample probability distribution 
$P(w^2)$.  
In addition, the average structure factor associated to the
critical configuration is
\begin{equation}
\label{eq:def-sdeq}
S_q = \frac{1}{L} \overline{ \left| \sum_{j=0}^{L-1} u^c_j \, e^{-iqj} \right|^2
 },
\end{equation}
where $q=2\pi n/L$, with $n=1,...,L-1$. One can show, using a simple dimensional
analysis, that given a roughness exponent $\zeta$, such that $\overline{w^2}
\sim L^{2 \zeta}$, the structure factor behaves as $S(q) \sim q^{-(1+2\zeta)}$ 
for small $q$, thus yielding an estimate to $\zeta$ without changing $L$.
To compute averages over disorder and sample-to-sample fluctuations, we consider 
between $10^3$ and $10^4$ independent disorder realizations depending on the size of the system.

\section{Results}
\subsection{Roughness at the critical point}
\label{s:exponents}

\begin{figure}[!tbp]
\begin{center}
\includegraphics[width=0.45\textwidth]{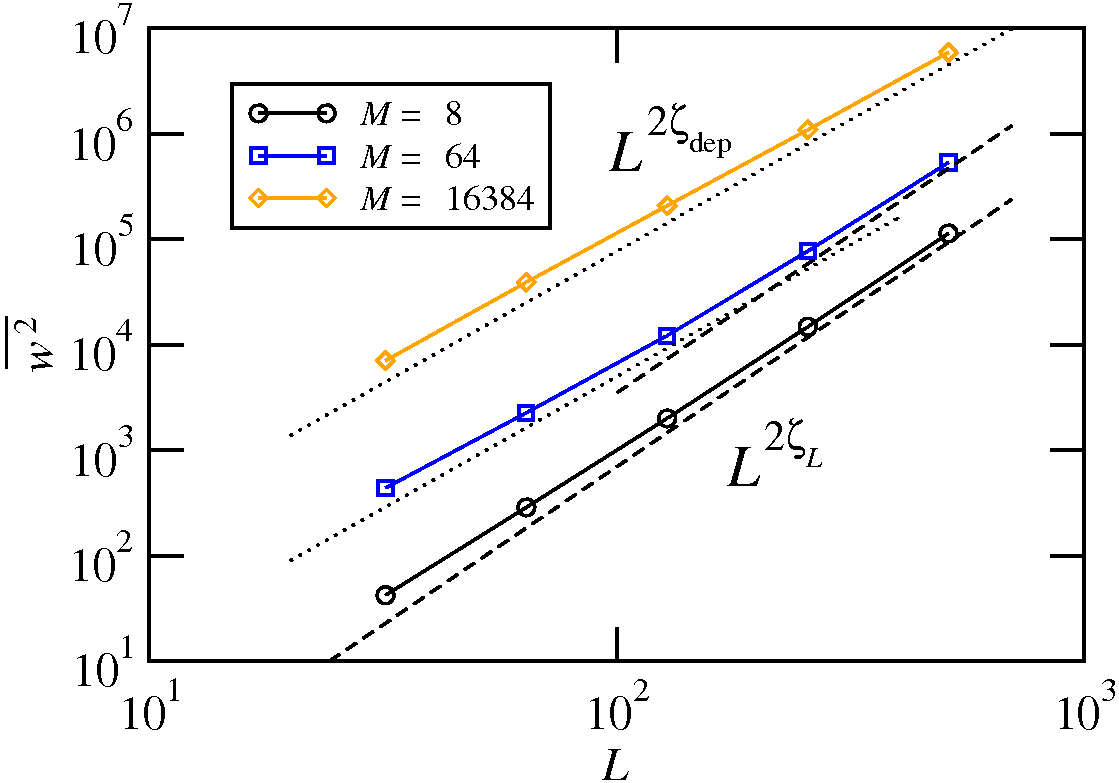}
\end{center}
\caption{\label{f:w2deL-Ap1} The scaling of $\overline{w^2}$ for the critical
configuration at different $M$ values as indicated. The curves for $M = 64$ and
$16384$ are shifted upwards for clarity. The dashed and dotted lines are guides
to the eye showing the expected slopes corresponding to the different roughness
exponents.}
\end{figure}

Figure~\ref{f:w2deL-Ap1} shows the scaling of the square width of the critical configuration 
$\overline{w^2}$ with the longitudinal size of the system $L$ for
$L=32, 64, 128, 256, 512$ and different values of $M$. 
When $M$ is small, $M=8$, 
for all the $L$ values shown we observe $\overline{w^2} \sim L^{2\zeta_{\mathrm{L}}}$ 
with $\zeta_{\mathrm{L}}=1.5$, corresponding to the Larkin exponent in $(1+1)$ dimensions. 
This value is different from the value $\zeta_{\mathrm{dep}}=1.25$~\cite{duemmer2,rosso_roughness_at_depinning}
expected for the random-manifold universality class, and is thus indicating that the periodicity effects 
are important for this joint values of $M$ and $L$. This situation is schematically represented in Fig.~\ref{fig:schema}(b). 
This result is a numerical confirmation of the two-loop functional renormalization 
group result of Ref.~\cite{ledoussal_frg_twoloops} which shows that 
the $\zeta=0$ fixed point, leading to a universal logarithmic growth of displacements 
at equilibrium is unstable. The fluctuations are governed, instead, by a coarse-grained generated 
random-force as in the Larkin model, yielding a roughness exponent $\zeta_{\mathrm{L}}=(4-d)/2$ in $d$
dimensions~\cite{ledoussal_frg_twoloops}, which agrees with our result for $d=1$.
We can thus say that for small enough $M$ (compared to $L$) the system belongs 
to the same random-periodic depinning universality class as charge density 
wave systems~\cite{narayan_fisher_cdw,narayan_fisher_depinning}, which strictly 
correspond to $M=1$.

When $M$ is large, on the other hand, $M=16384$ in Fig.~\ref{f:w2deL-Ap1}, for all the $L$ values considered the exponent 
is consistent with $\zeta_{\mathrm{dep}}$, of the random-manifold universality class. This 
situation is schematically represented in Fig.~\ref{fig:schema}(c), and we will show later 
that, for this elongated samples, the effects of extreme value statistics are already visible.

For intermediate values of $M$, such as $M=64$ in Fig.~\ref{f:w2deL-Ap1}, we 
can observe the crossover in the scale-dependent roughness exponent 
$\zeta(L) \sim \frac{1}{2} \frac{d\log w^2}{d\log L}$ 
changing from $\zeta_{\mathrm{dep}}$ to
$\zeta_{\mathrm{L}}$ as $L$ increases, as indicated by the dashed and dotted lines. 
This crossover, from the random-manifold to the random-periodic depinning geometry, 
occurs at a characteristic distance $l^* \sim M^{1/\zeta_{\mathrm{dep}}}$, when 
the width in the random-manifold regime reaches the transverse dimension or periodicity $M$. 
At finite velocity, this crossover length remains constant up to a non-trivial characteristic 
velocity and then decreases with increasing velocity~\cite{bustingorry_RM-RP}.

\begin{figure}[!tbp]
\begin{center}
\includegraphics[width=0.45\textwidth]{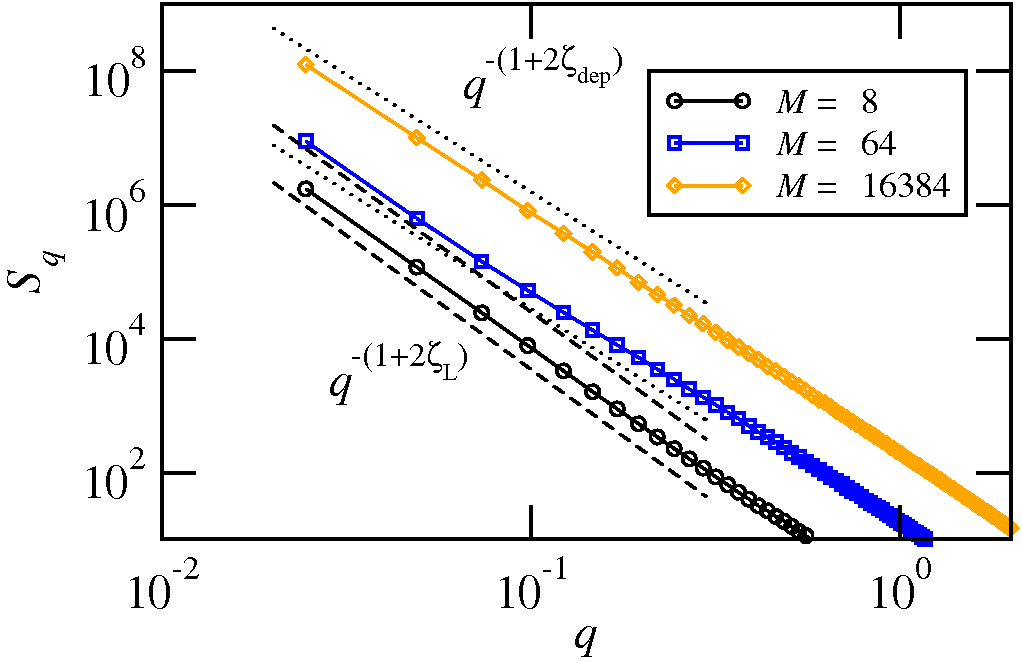}
\end{center}
\caption{\label{f:fcyw2_L256-Ap1} Structure factor of the critical configuration
for $L=256$ and different $M$ values, as indicated. The curves for $M = 64$ and
$16384$ are shifted upwards for clarity. The dashed and dotted lines are guides
to the eye showing the expected slopes corresponding to the different roughness
exponents.}
\end{figure}

\begin{figure}[!tbp]
\begin{center}
\includegraphics[width=0.45\textwidth]{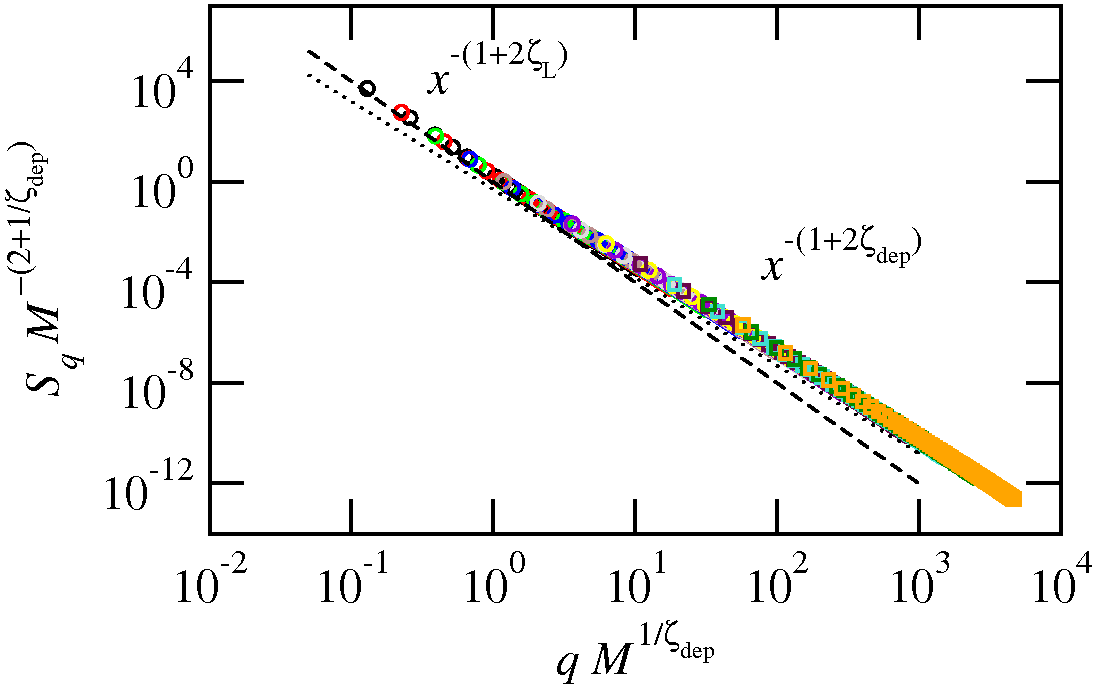}
\end{center}
\caption{\label{f:fcyw2_L256-Ap1-scal} Scaling of the structure factor of the
critical configuration for $L=256$ and different values of the transverse size
$M=2^p$ with $p=3,4,...,14$ $M$. Although the values of the two exponents are
very close, the change in the slope of the scaling function against the scaling
variable $x = q \, M^{1/\zeta_{\mathrm{dep}}}$ is clearly observed.}
\end{figure}

The above mentioned geometrical crossover can be studied in more details through 
the analysis of the structure factor $S(q)$, for a line of fixed size $L$. 
In Fig.~\ref{f:fcyw2_L256-Ap1} we show $S(q)$ for $L=256$ and $M=8,64,16384$. 
For the intermediate value $M=64$ a crossover between the two regimes is
visible, and can be described by
\begin{equation}
 S_q \sim
\left\{
\begin{array}{ll}
 q^{-(1+2\zeta_{\mathrm{L}})} & q \ll q^*, \\
 q^{-(1+2\zeta_{\mathrm{dep}})} & q \gg q^*. \\
\end{array}
\right.
\end{equation}
with $q^*$ expected to scale as $q^* \sim l^{*-1} \sim M^{-1/\zeta_{\mathrm{dep}}}$.
Therefore, the structure factor should scale as $S_q
M^{-(2+1/\zeta_{\mathrm{dep}})} = H(x)$, where the scaled variable is $x = q \,
M^{1/\zeta_{\mathrm{dep}}} \sim q/q^*$ and the scaling function behaves as
\begin{equation}
H(x) \sim \left\{
\begin{array}{ll}
 x^{-(1+2\zeta_{\mathrm{L}})} & x \ll 1, \\
 x^{-(1+2\zeta_{\mathrm{dep}})} & x \gg 1. \\
\end{array}
\right.
\end{equation}
The collapse of Fig.~\ref{f:fcyw2_L256-Ap1-scal} for $L=256$ and
different values of $M=2^p$ with $p=3,4,...,14$ shows that this 
scaling form is a very good approximation. However, as we show 
below, small corrections can be expected fully in the random-manifold regime
in the large $M L^{-\zeta_{\mathrm{dep}}}$ limit of very elongated samples.

\begin{figure}[!tbp]
\begin{center}
\includegraphics[width=0.45\textwidth]{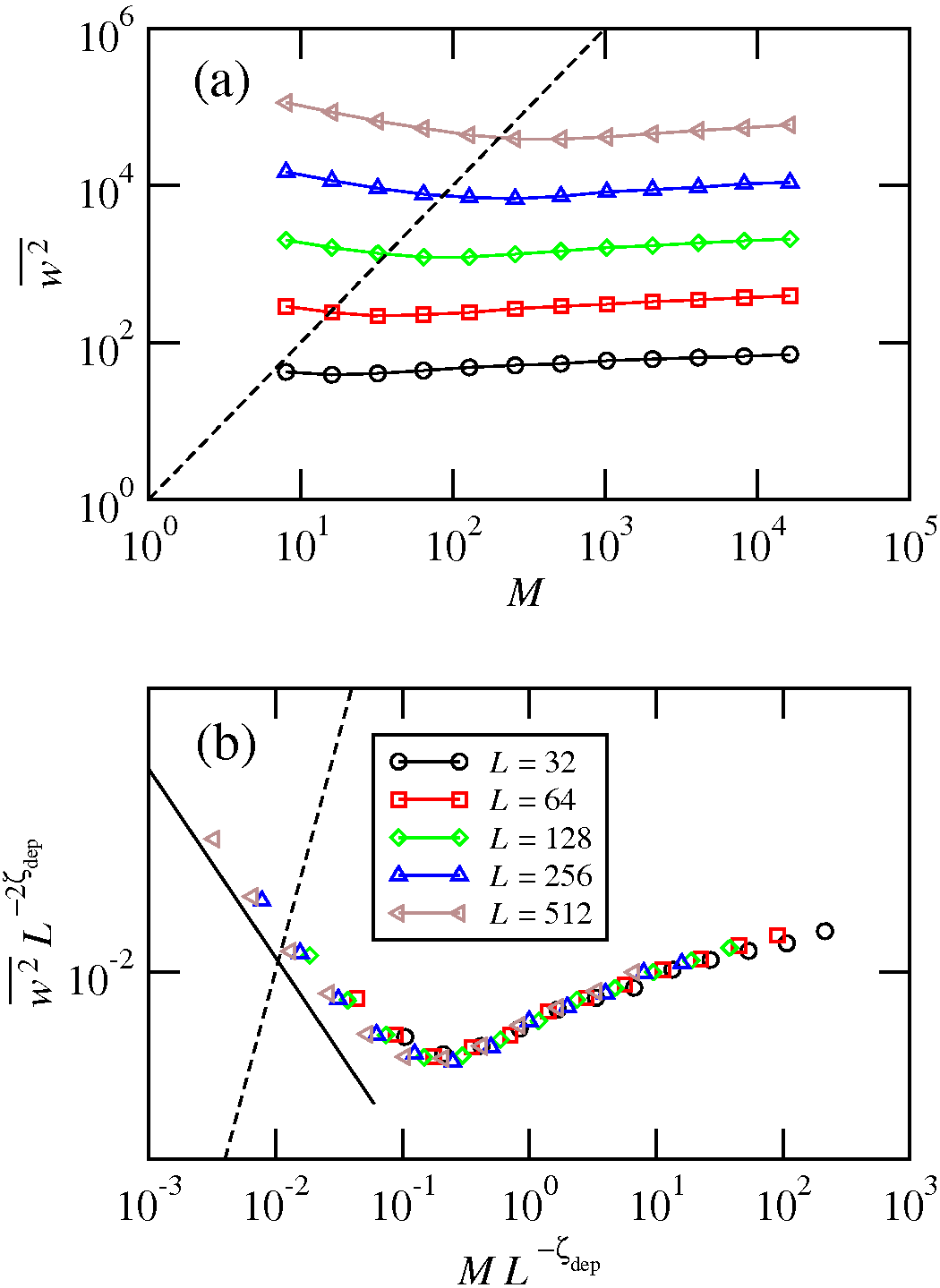}
\end{center}
\caption{\label{f:w2deM-D1-scal} (a) Squared width of the critical configuration
as a function of $M$ for different system sizes $L$ as indicated. (b) Scaling of
the width in (a), showing that the relevant control parameter is
$M/L^{\zeta_{\mathrm{dep}}}$.  The dashed line in (a) and (b) corresponds to
$\overline{w^2} = M^2$, which is always  to the left of the minimum of
$\overline{w^2}$ occurring at $k^*=m^* L^{-\zeta_{\mathrm{dep}}}$. The solid line indicates 
$k^{2(1-\zeta_{\mathrm{L}}/\zeta_{\mathrm{dep}})}$
which is the behavior expected purely from the random-periodic to random-manifold 
crossover at the characteristic distance $l^*\sim M^{1/\zeta_{\mathrm{dep}}}$.}
\end{figure}

In Fig.~\ref{f:w2deM-D1-scal}(a), we show $\overline{w^2}$ as a function of 
the transverse periodicity $M$ for different values of the longitudinal periodicity 
$L$. Remarkably, $\overline{w^2}$ is a non-monotonic function of $M$. For small $M$ 
it decreases towards an $L$ dependent minimum $m^*$, 
and then increases with increasing $M$, in the regime where the extreme value statistics 
starts to affect the distribution of the critical force~\cite{bolech_critical_force_distribution}. 
Since the only typical transverse scale in Fig.~\ref{f:w2deM-D1-scal}(a) is set by 
the minimum $m^*$, we can expect  $\overline{w^2} \sim m^{*2} G(M/m^*)$ with $G(x)$ 
some universal function. On the other hand, since the only relevant characteristic 
length-scale of the problem is set by the crossover between the 
random-periodic regime and the random-manifold regime, we can simply 
write $m^* \sim L^{\zeta_{\mathrm{dep}}}$ and therefore
\begin{equation}
\label{eq:w2G}
 \overline{w^2} \, L^{-2 \zeta_{\mathrm{dep}}} \sim G(M
\,L^{-\zeta_{\mathrm{dep}}}).
\end{equation}
This scaling form is confirmed in Fig.~\ref{f:w2deM-D1-scal}(b)
and shows that the aspect-ratio parameter 
$k=M L^{-\zeta_{\mathrm{dep}}}$ fully controls the 
anisotropic finite-size scaling of the problem. 
It is worth, however, noting some interesting 
consequences of the result of Fig.~\ref{f:w2deM-D1-scal}(b), 
as we describe below.

Since at very small $k$ the interface is in the random-periodic 
regime, Eq.~(\ref{eq:w2G}) should led to $\overline{w^2} \sim L^{2 \zeta_{\mathrm{L}}}$ and therefore one deduces that,
\begin{equation}
\label{eq:Gsmallk}
G(k)\sim k^{2(1-{\zeta_{\mathrm{L}}}/{\zeta_{\mathrm{dep}}})},\;\;\;  k \ll k^*,
\end{equation}
where $k^*=m^* L^{-\zeta_{\mathrm{dep}}}$.
The fact that the random-periodic roughness exponent ${\zeta_{\mathrm{L}}} =  3/2$  
is larger than the random-manifold one ${\zeta_{\mathrm{dep}}}\approx 5/4$ 
consequently implies an initial decrease of $G(k)$ as $G(k)\sim k^{-2/5}$, as shown 
in Fig.~\ref{f:w2deM-D1-scal}(b) by the solid line. Periodicity effects, or the crossover from random-periodic to random-manifold, 
thus explain the initial decrease of $G(k)$ seen in Fig.~\ref{f:w2deM-D1-scal}(b), 
or the initial decrease of $\overline{w^2}$ against $M$ for fixed $L$, 
seen in Fig.~\ref{f:w2deM-D1-scal}(a). 
At this respect, it is then worth noting that the line $\overline{w^2} = M^2$, 
shown by a dashed line, lies completely in the regime 
$k < k^*$ implying that the naive criterion $\overline{w^2} < M^2$ is not enough
to avoid periodicity effects, and to have the system fully in the random-manifold regime.
As we show later, this is related with the shape of the probability 
distribution of $P(w^2)$ which displays sample-to-sample fluctuations of 
the order of the average $\overline{w^2}$.

The presence of a minimum at $k^*$ in the function $G(k)$ and in particular 
its slower than power-law increase for $k>k^*$ is non-trivial and constitutes 
one of the main results of the present work. This result shows 
that corrections to the standard scaling $\overline{w^2}\sim L^{\zeta_{\mathrm{dep}}}$ 
may arise from the aspect-ratio 
dependence of the prefactor $G(k)$. On the one hand, $\overline{w^2}$ now grows with $M$ for $L$ 
fixed, in spite that $\overline{w^2} \ll M^2$, i.e. transverse-size/periodicity scaling 
is present. On the other hand, the scaling of $\overline{w^2}$ with $L$ is slower in this regime, 
due to subleading scaling corrections coming from $G(k)$. The precise origin 
of these interesting leading and subleading corrections in the finite-size 
anisotropic scaling are highly non-trivial. 
Since the critical 
configurations in this regime have the constant 
roughness exponent $\zeta_{\mathrm{dep}}$ of the random-manifold 
universality class, the slow increase of $G(k)$ cannot be attributed 
to a geometrical crossover effect, as for the case $k<k^*$. 
However, we might relate this effect to the crossover in 
the critical force statistics, from Gaussian 
to Gumbel, in the $k \gg k^*$
limit~\cite{bolech_critical_force_distribution}. In the Gumbel regime, 
the average critical force is expected to increase 
as $F_c \sim \log (M/L^\zeta_{\mathrm{dep}}) \equiv \log k$~\cite{fedorenko_frg_fc_fluctuations}, 
since the sample critical force can 
be roughly regarded as the maximum among $M/L^\zeta_{\mathrm{dep}}$ independent 
sub-critical forces and configurations~\cite{bolech_critical_force_distribution}. 
The increase in the critical force might be therefore correlated with 
the slow increase of roughness. The physical connection between the two is 
subtle though, since a large critical force in a very elongated sample 
could be achieved both by profiting very rare correlated pinning forces 
such as accidental columnar defects, or by profiting very rare 
non-correlated strong pinning forces. Since in the first case 
the critical configuration would be more correlated and in general less rough than for 
less elongated samples (smaller $k$), contrary to our numerical data  of Fig.~\ref{f:w2deM-D1-scal}(b), 
we think that the second cause is more plausible. We can thus think that 
in the $k \gg k^*$ limit of extreme value statistics of $F_c$, the effective disorder 
strength on the critical configuration increases with $k$. This might be translated  
into the universal function $G(k)$, such that 
$\overline{w^2} \approx L^{2 \zeta_{\mathrm{dep}}} G(k)$ can increase 
for increasing values of $k$ at fixed $L$ in such regime. 
A quantitative description of these scaling corrections remains 
an open challenge.

\subsection{Distribution function}
\label{s:distributions}

\begin{figure}[!tbp]
\begin{center}
\includegraphics[width=0.45\textwidth]{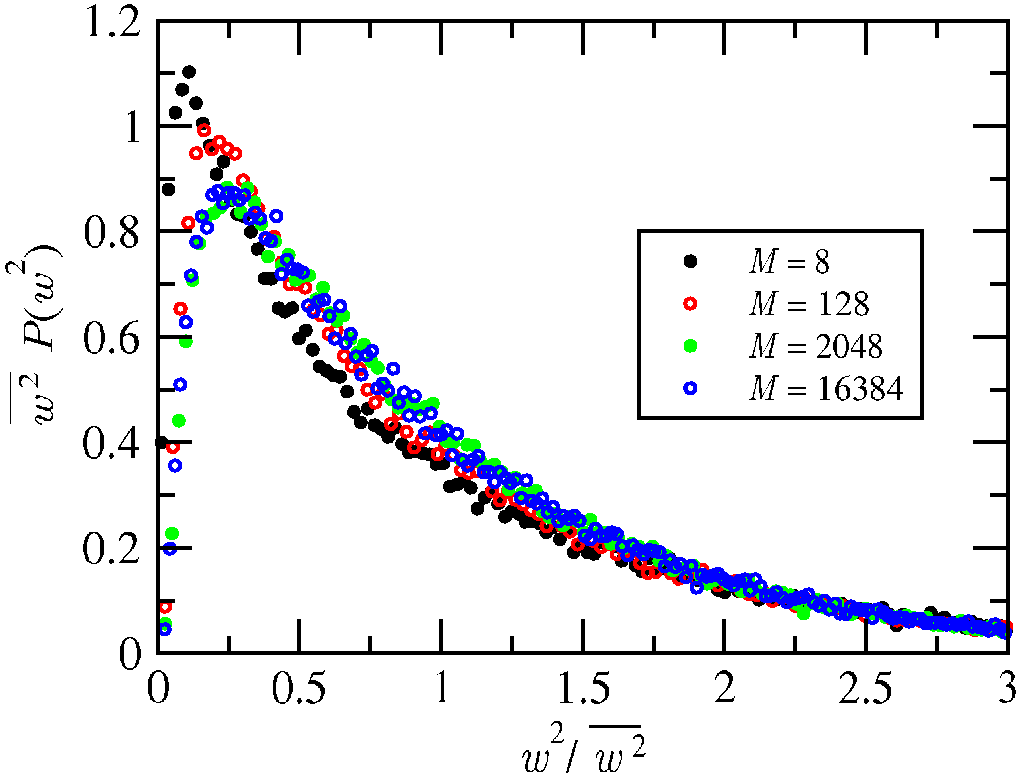}
\end{center}
\caption{\label{f:PhideM-L256} Scaling function $\Phi(x)$ for $L=256$ and
different values of $M=8,128,2048,16384$, which shows the change with the
transverse size $M$.}
\end{figure}

We now analyze sample-to-sample fluctuations of the square width $w^2$ by computing 
its probability distribution $P(w^2)$. This property is relevant as $w^2$ fluctuates 
even in the thermodynamic limit for critical interfaces with a 
positive roughness exponent~\cite{Racz2003}. It has been 
computed for models with dynamical disorder  
such as random-walk~\cite{Foltin1994} or Edwards--Wilkinson interfaces~\cite{Antal1996,Bustingorry2007}, 
the Mullins Herrings model~\cite{Plischke1994} and for non-Markovian Gaussian signals 
in general~\cite{rosso_santachiara_2005,santachiara_rosso_2007}. It has also been 
calculated for non-linear models such as the one-dimensional Kardar--Parisi--Zhang 
model~\cite{Marinari2002,Bustingorry2007a} and for the 
quenched Edwards--Wilkinson model at equilibrium~\cite{Bustingorry2006}.

In particular, the probability distribution $P(w^2)$ of critical 
interfaces at the depinning transition  was studied analytically~\cite{ledoussal_wiese_pdf2003}, 
numerically~\cite{rosso_width_distribution} and also 
experimentally for contact lines in partial wetting~\cite{moulinet_distribution_width_contact_line2}. 
Remarkably, non-Gaussian effects in depinning models are found to be smaller 
than $0.1\%$~\cite{rosso_width_distribution,ledoussal_wiese_pdf2003}, thus showing 
that $P(w^2)$ is strongly determined by the self-affine (critical) geometry itself, rather 
than by the particular mechanism producing it.  
As in all the above mentioned systems the width distribution $P(w^2)$ at different universality classes 
of the depinning transition was found to scale as
\begin{equation}
\overline{w^2} P(w^2) \approx \Phi_{\zeta} \left( \frac{w^2}{\overline{w^2}}
\right).
\end{equation}
with $\Phi_{\zeta}$ an universal function, which only depends on the roughness exponent 
$\zeta$ and on boundary conditions when the global width is 
considered~\cite{rosso_santachiara_2005,santachiara_rosso_2007}.
In this way, $\overline{w^2}$ is the only characteristic 
length-scale of the system, absorbing 
the system longitudinal size $L$, and all the non-universal parameters of the model such as 
the elastic constant of the interface, the strength of the disorder and/or the temperature. 
Since $\Phi_{\zeta}$ can be easily generated using non-Markovian Gaussian 
signals~\cite{krauth_smac}, the quantity $\overline{w^2} P(w^2)$ is a good observable 
to extract the roughness exponent of a critical interface from experimental data. 

\begin{figure}[!tbp]
\begin{center}
\includegraphics[width=0.45\textwidth]{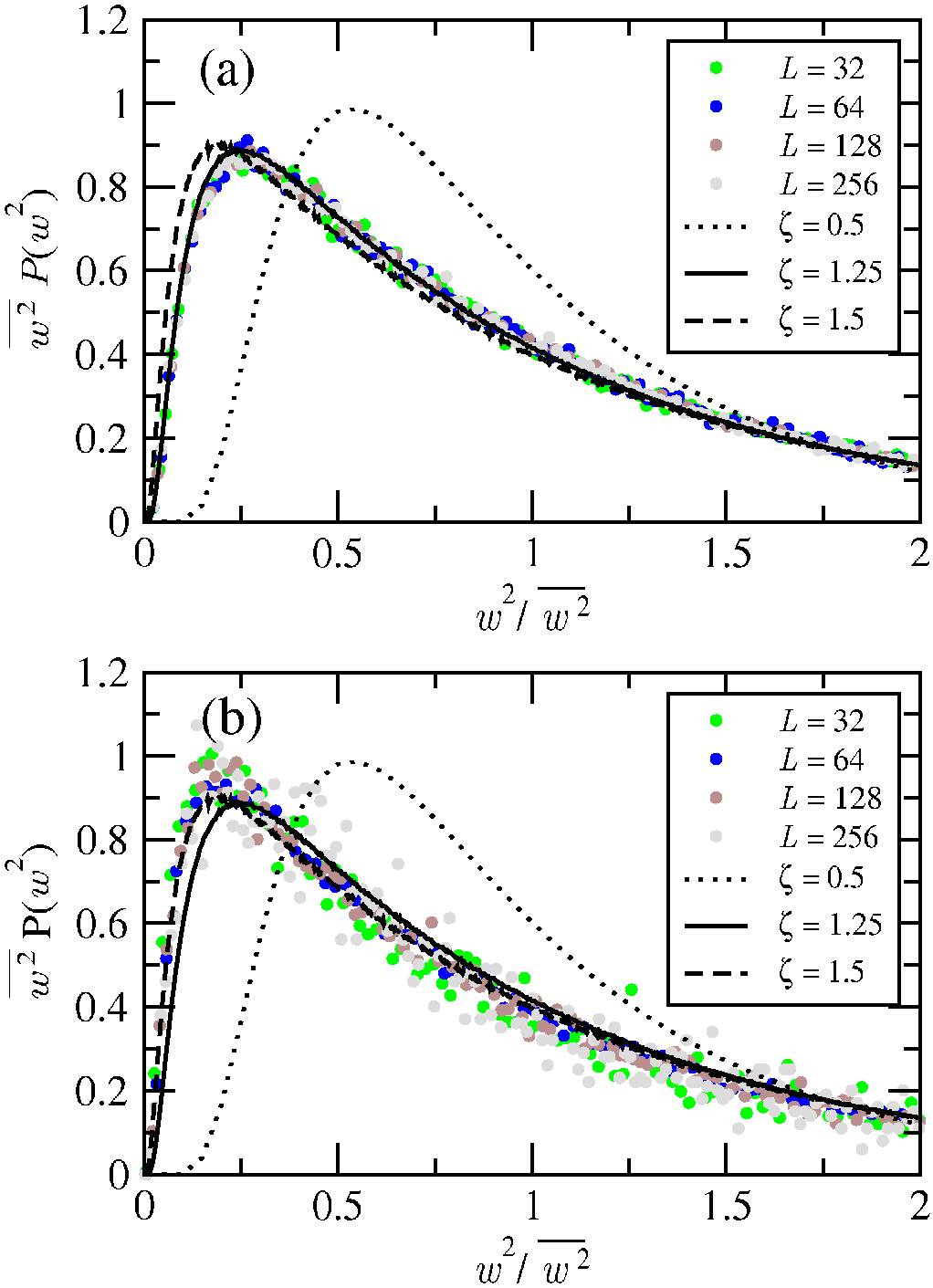}
\end{center}
\caption{\label{f:PhideM-scal} Scaling function $\Phi(x)$ for different values
of $L=32,64,128,256$ while keeping (a) $k=M/L^{\zeta_{\mathrm{dep}}} \approx 1$
and (b) $k=M/L^{\zeta_{\mathrm{dep}}} \approx 0.025$. The dotted line corresponds
to the scaling function of the non-disordered Edwards--Wilkinson
equation~\cite{Foltin1994}, while the continuous and dashed lines correspond to
the scaling functions of Gaussian signals with $\zeta=1.25$ and $\zeta=1.5$,
respectively~\cite{rosso_width_distribution,krauth_smac}.}
\end{figure}

In Fig.~\ref{f:PhideM-L256}, we show how the scaled distribution function
$\Phi(x) \equiv \overline{w^2} \, P(x \;\overline{w^2})$ 
looks like for the depinning transition 
in a random-periodic medium for a fixed value $L=256$ and different values of $M$.
We see that $\Phi(x)$ depends on $M$ for small $M$ but converges to a 
fixed shape for large $M$. We also note that for all $M$ 
$\Phi(x)$ extends appreciably beyond $x=1$ explaining why the 
criterion $\overline{w^2} \lesssim M^2$ is not enough to be fully 
in the random-manifold regime, as noted 
in Fig.~\ref{f:w2deM-D1-scal}. 

In Fig.~\ref{f:PhideM-scal}, we show the scaling function $\Phi(x)$
for different values of $L$ and $M$ but fixing the aspect-ratio parameter
$k=M/L^{\zeta_{\mathrm{dep}}}$,  $k \approx 1 > k^*$ in Fig.~\ref{f:PhideM-scal}(a) 
and $k \approx 0.025 \ll k^*$ in Fig.~\ref{f:PhideM-scal}(b), with $k^*$ the minimum 
of $\overline{w^2}$. Since 
data for the same $k$ practically collapses into the same curve, we can write for our case:
\begin{equation}
\label{eq:PhideM-scal}
\overline{w^2} P(w^2) = \Phi\left( \frac{w^2}{\overline{w^2}},k \right).
\end{equation}
Therefore, the anisotropic scaling of the probability distribution is fully 
controlled by $k$, as it was found for $\overline{w^2}$.

In Figs.~\ref{f:PhideM-scal}(a) and (b), we also show 
the universal functions $\Phi_{\zeta_{L}}$ and 
$\Phi_{\zeta_{\mathrm{dep}}}$ generated using non-Markovian 
Gaussian signals~\cite{rosso_width_distribution,krauth_smac}, and for comparison 
we also show $\Phi_{1/2}$ corresponding to the Markovian periodic 
Gaussian signal or the Edwards--Wilkinson equation~\cite{Foltin1994}. Comparing this with the collapsed data for depinning, we see 
that the function $\Phi\left( \frac{w^2}{\overline{w^2}},k \right)$ respects 
the limits
\begin{eqnarray}
\Phi\left( x,k \to 0 \right) &=& \Phi_{\zeta_{\mathrm{L}}}(x), \nonumber \\
\Phi\left( x,k \gtrsim k^* \right) &\approx& \Phi_{\zeta_{\mathrm{dep}}}(x),
\end{eqnarray}
as expected from the existence of the geometric crossover between 
the roughness exponents ${\zeta_{\mathrm{L}}}$ for 
$k\to 0$ and ${\zeta_{\mathrm{dep}}}$ for $k > k^*$.
For intermediate values $k<k^*$, however, 
$\Phi\left( \frac{w^2}{\overline{w^2}},k \right)$ does not 
necessarily coincide with the one of a Gaussian signal function $\Phi_{\zeta}$ 
for a given $\zeta$, since the critical configuration includes a 
crossover length $l^* \lesssim L$. Whether multi-affine 
or effective exponent self-affine non-Markovian Gaussian 
signals can be used to describe satisfactorily 
these intermediate cases is an interesting open issue.

\section{Conclusions}
\label{s:discussion}
We have numerically studied the anisotropic finite-size scaling 
of the roughness of a driven elastic string at its sample-dependent 
depinning threshold in a random medium with periodic boundary conditions 
in both the longitudinal and transverse directions. 
The average square width $\overline{w^2}$ and its   
probability distribution are both controlled by the  
parameter $k=M/L^{\zeta_{\mathrm{dep}}}$.
A non-trivial single minimum for a finite
value of $k$ was found in $\overline{w^2}/L^{2\zeta_{\mathrm{dep}}}$. 
For small $k$, the initial decrease of $\overline{w^2}$ reflects the crossover 
from the random-periodic 
to the random-manifold roughness. 
For very large $k$, the growth with $k$ implies that the crossover to 
Gumbel statistics in the critical forces induces corrections to $G(k)$, that grow 
with $k$, to the string roughness scaling 
$\overline{w^2}\approx G(k) L^{2\zeta_{\mathrm{dep}}}$. These increasingly rare 
critical configurations thus have an anomalous roughness scaling: they have a 
transverse-size/periodicity scaling in spite that its width is $\overline{w^2} \ll M^2$,
and subleading (negative) corrections to the standard random-manifold longitudinal-size scaling.

Our results could be useful for understanding roughness fluctuations and scaling
in finite experimental systems. The crossover from random-periodic to random-manifold 
roughness could be studied in periodic elastic systems with variable periodicity, 
such as confined vortex 
rows~\cite{kokubo_vortex_rows2004} and single-files of macroscopically 
charged particles~\cite{coste_single_file_diffusion2010} or colloids~\cite{herrera_single_file_colloids},
with additional quenched disorder. The rare-event dominated scaling corrections to 
the interface roughness scaling could be studied in systems with a large transverse dimension, 
such as domain walls in ferromagnetic nanowires~\cite{kim_dimensional_crossover_creep}.
For the later case, it would be interesting to have a quantitative theory, 
making the connection between the extreme value statistics of the depinning threshold 
and the anomalous scaling corrections to the roughness of such rare 
critical configurations. This would allow to understand the dimensional 
crossover, from interface to particle depinning.

\begin{acknowledgements}
This work was supported by CNEA, CONICET under Grant No. PIP11220090100051͒, and ANPCYT under Grant No. PICT2007886. A. B. K. acknowledges Universidad de Barcelona, Ministerio de Ciencia e
Innovaci\'{o}n (Spain) and Generalitat de Catalunya for partial funding through
I3 program.
\end{acknowledgements}



\end{document}